**Evaluating the Efficacy of Online Assessments in Higher Education**
Author : Santhosh Pogaku

**Abstract**:
This experiment research study examines how traditional assessment methods, such as written tests and presentations, compared to the new online tests in higher education. We want to know how the use of the Internet for assessment affects how well students do and what they learn. Custom tests are individualised, just like your regular tests. On the other hand, online testing allows students to take the test from anywhere using the internet. We're trying to determine if these student assessment changes make a difference. There are some good things about online testing. Students are given greater flexibility because they can take the exam from home or anywhere they like. This can help a variety of students. But there are also challenges, such as technology problems and the need for students to be proficient in the use of technology. We will compare how students do on traditional tests versus online tests. We will examine the statistics and hear what students say about their experiences. We will consider where students come from, how comfortable they are with technology, and what they want for learning. By conducting this research, we hope to help teachers and schools understand how to assess students in this technological age and determine what works best for everyone.

**Introduction**
In this research study, we examine the dynamics of traditional assessment methods and the growth of online assessment in higher education. Our approach, which focuses on revealing subtle effects on student performance and learning outcomes, is based on an experimental research design. Through a systematic review, we aim to understand the implications of the transition from traditional assessment methods to online assessment in the higher education context. Our study adopts a comprehensive approach, using both qualitative and quantitative research to capture the full meaning of multidimensional influences on students. By carefully designing experimental and control groups, we seek to identify potential changes in student performance and learning outcomes within specific assessment strategies. This research is not just an academic exercise; It is a dedicated effort to inform scholars, institutions, and policymakers about the complex dynamics at play. As we move into uncharted territories of educational research, our study aspires to provide valuable insights that will shape the future state of research practices in higher education. This study exam lines how well online assessment in professional mathematics compares with traditional online assessment in measuring students' learning. We use data from mathematics courses taken at our university to see if online tests measure the same as in-person tests. Interestingly, we find that students tend to score higher on online tests, possibly due to factors such as computer anxiety and students' perceptions of online research in mathematics classes [1]. The impact of the Internet on education, especially online learning, has increased worldwide, especially during the COVID-19 pandemic. This study examines the usefulness, reliability, safety, and psychological aspects of student perceptions of online multiple-choice testing. The findings with 127 business college students provide valuable insights for future applications [2]. This study examines the impact of online time management on student performance in computer science courses, as well as time management preferences and outcomes. The one-semester survey with 113 participants shows that students enjoy visual countdown moments, especially those with colour changes. Interestingly, time students' time choices did not significantly affect their performance [3]. The global impact of the COVID-19 pandemic on education has accelerated the need to turn to virtual learning, leading to increased reliance on online assessment tools. This study focuses on developing a flexible e-assessment system in response to the current educational climate, emphasising efficiency and reducing plagiarism. Statistical methods are used to analyse the performance of engineering college students during this period of change [4]. Addressing the limitations of traditional e-assessment in assessing high-level technical students' skills, this paper uses askMe! System—an innovative, personalised, and interactive e-assessment solution. It emphasises the program's features and role in acquiring general knowledge, and the analysis concludes with a dual perspective examining the authors and the students [5]. Protecting academic integrity is paramount due to the unique challenges of detecting and preventing cheating in online education. This paper presents object recognition methods for identifying work patterns in students' timelines in online learning programs. Teachers can locate unauthorised collaborative projects by analysing detailed timelines, such as collaborating on experiments or creating external copies. Based on years of implementation in an online career-based course, this approach proved effective in potential fraud cases [6]. Assessment is essential for interpreting student performance, guiding decision-making for various stakeholders, and assessing skills. Often, assessment is conflated; assessment focuses on evaluating subject performance and measuring performance, while assessment judges students against criteria. This paper proposes an M-assessment, which often conducts immediate research on practice tests and thematic questions for advanced courses [7]. This paper presents a new framework for addressing freezes in online learning research, especially regarding physical requirements for proctored tests. The proposed model enables remote and electronic monitoring through explicit authentication, ensures non-interference, and maintains identity integrity in formal research. Computational analysis of the model confirms the feasibility of this approach [8]. In light of the challenges posed by the COVID-19 pandemic, this study explores the perspectives of students and faculty on the effectiveness of the Multiple Attempt Format (MAF) in online assessment and its role in addressing academic integrity concerns. Research on perceptions of the MAF and the Academic Honesty Questionnaire (PMAHQ) shows that students strongly agree (89%) that the MAF adequately addresses issues of academic honesty. Teachers (100%) also express positive attitudes toward MAF, highlighting its role in teaching responsibilities and enabling students to make informed decisions to improve grades [9]. This review explores the diverse landscapes of online assessment applications from Windows-based to web-based, licensed to free software, to proprietary standardised. These applications include critical assessment steps, including materials, training, assessment, and awarding databases with detailed data stored in. The proposed methodology recommends data mining to detect and prevent computer fraud and trace students' actions, IP addresses, and operational data policies [9].

RQ: Compared to traditional assessment techniques, how do online evaluations affect higher education student performance and learning outcomes?

**Methodology**

**Sample**
Each group had a maximum of ten students, excluding leaders. We used the random sampling method to generate these groups. In Excel, we used the rand() function to assign random numbers to each sample and sorted them in ascending order. The first six sample became the experimental group, and the next five formed the control group. This split was done randomly, ensuring that each group member had an equal chance of joining each group. Sample in each group will use a set of assessments to present to their partners.

**Measure**
This research method is based on the "behavioural Intention" model described in reference [11]. The central assumption of this study is that the more frequently individuals use a device, the greater their proficiency. Although the survey results were also subjected to statistical analysis, including means, medians, standard deviations, and standard deviations, the leading indicators were derived from the survey results on a Likert scale. A scale from 1 to 7 assessed the information, yielding a comprehensive rating.

**Design**
The structure of this experimental research concerned splitting the allocated groups into two teams assigned to the research group and another to the control group. Google Forms facilitated the advent of these surveys. A summary of every sample was submitted to ensure randomness and eliminate bias, after which the complete set was randomised. The first five samples from the rearranged list have been assigned to the experimental group, even though the last five were assigned to the control group. We plan to conduct four surveys for the experiment sample and identical for the control sample. Program experts have tested each survey template, and we've got advanced versions to ensure accurate results.

**Procedure**
As part of the survey, we crafted a complex list of questions, and the experimental group received an introductory video about the research observation. Following this, the experimental group completed a survey, and we awaited their responses. Participants have been given an appropriate timeframe to finish the study, with notifications based on their availability. In comparison, the control group received equal surveys for amassing responses but were not admitted to any precise videos or sample facts.

As the undertaking approached its conclusion on November 29th, 2023, survey bureaucracy was sent out, and responses began coming in on an equal date. The survey included various Home Automation-associated subjects, aiming to accumulate qualitative facts. The statistics obtained through this technique have been cautiously considered and integrated to yield the most effective results for statistical evaluation.

To check the respondents within the test, we used the BI model. Information exchange commenced handiest after each group had submitted all their survey responses. The following figure illustrates the phased method of this test. In the formation of deterministic representations, the 'R' (random sample), 'X' (technique of treatment), and 'O' (commentary) factors have been considered for the post-test phase. The actual framework for the publish-take was chosen randomly, as told with the professor's aid and money owed for the existence of companies.

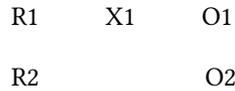

In the 2x2 factor diagram above, we have R1 representing the random sample of the experimental group, R2 representing the random sample of the control group, X1 representing the verbal or video test used for the experimental group, and O1 describing the observation.

In addition to asking standard questions about name, gender, and age, surveys included measurement questions and questions about home automation. The findings were used as discrete variables for statistical analysis.

Even using a random sampling procedure to create the research group, some participants may need to be adequately informed about the subject under study, which may introduce errors. Participants were appreciated for their involvement, valuable time, and expertise in this study.

**Results**

The data obtained is based on the 'Behavioral Intention' model, and the outcomes that form the foundation of this framework have been considered in the table below. The evaluation factors labelled BI1, BI2, BI3, BI4, and BI5 represent behavioural intention estimated from responses from various questionnaires for samples S1, S2, S3, S4, and S5. Independent outcomes from the two groups were collected, and the resulting data was statistically analysed. The following table clarifies this. The responses were graded on a scale of 1 to 7, and the cumulative input indicates the degree of support for or opposition to the thesis study. The static diagram aided in the separation of results into two observations.

*Table 1: (Experimental Group)*

| BI | S1 | S2 | S3 | S4 | S5 |
|---|---|---|---|---|---|
| BI1 | 5 | 6 | 6 | 5 | 6 |
| BI2 | 5 | 4 | 5 | 7 | 7 |
| BI3 | 7 | 7 | 7 | 7 | 7 |
| BI4 | 6 | 6 | 6 | 6 | 4 |
| BI5 | 5 | 5 | 5 | 4 | 7 |
| Total | 28 | 28 | 29 | 29 | 31 |

*Table 2: (Control Group)*

| BI | S1 | S2 | S3 | S4 | S5 |
|---|---|---|---|---|---|
| BI1 | 5 | 6 | 5 | 5 | 6 |
| BI2 | 5 | 4 | 5 | 7 | 7 |
| BI3 | 6 | 5 | 6 | 7 | 6 |
| BI4 | 6 | 6 | 3 | 4 | 6 |
| BI5 | 5 | 5 | 5 | 4 | 5 |
| Total | 27 | 26 | 24 | 27 | 30 |

Based on the data and findings, examining the total calculated results, standard deviation, mean, median, deviation, and variance for the experimental and control groups, the mean observation of experimental group 1 is higher than the mean observation of control group 2. Neither the video nor the story was seen, and the following table shows the statistical analysis.

*Table 3: Statistic Analysis*

| | Mean | Variance | Standard Deviation | Median | MODE |
|---|---|---|---|---|---|
| Experiment Group | 29 | 1.6 | 1.2649 | 29 | 29 |
| Control Group | 26.8 | 3.2 | 1.7889 | 27 | 27 |

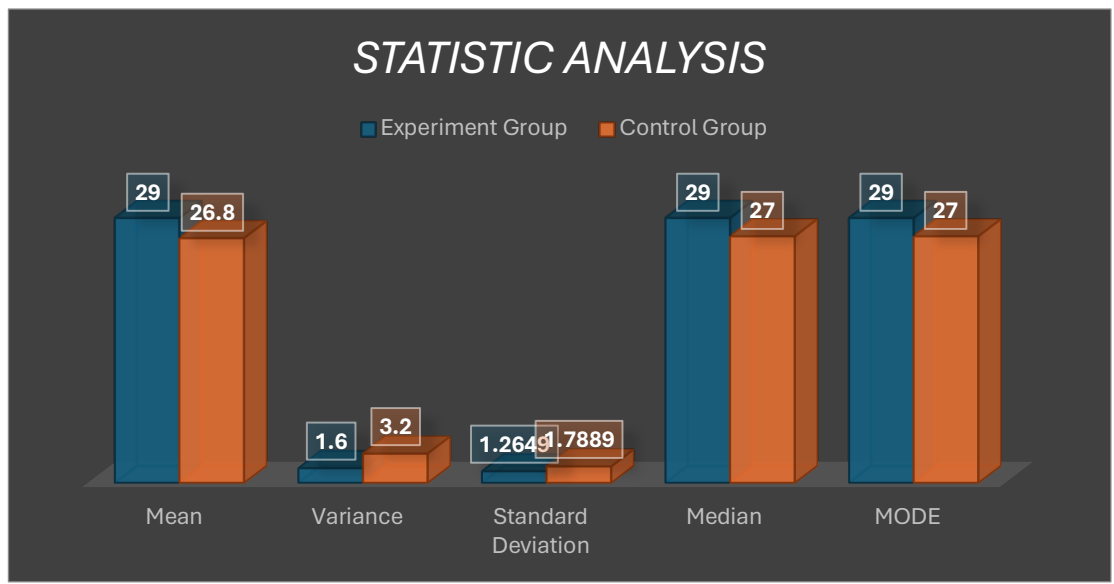

*Figure 1: Statistic Analysis*

Discussion

The results of the pre-post survey using a Likert scale (1 to 7) in both the experimental and control groups reveal remarkable insight into the impact of the intervention.

Pre-Survey comparisons:
Experimental group: The mean score of the experimental group was initially 1.6, generally indicating a positive attitude.
Control group: In contrast, the control group had a higher score of 3.2, indicating comparative neutrality.
The improvement in preassessment scores indicates that the control group may have held different perceptions or expectations than the experimental group before the intervention.

Post-Survey Impact:
Experimental group: After intervention, the experimental group showed little change in their mean scores, which increased slightly to 1.8. This suggests that the interventions may not have had an immediate and minimal effect on their mood.
Control group: Surprisingly, there was a significant change in the mean score for the control group, which decreased to 2.7. This unexpected variability raises questions about external factors or influences during the study period.

Comparative B/w Pre and Post:

The control group's higher preassessment scores and higher post-assessment scores may indicate the need for further investigation of possible external factors influencing their mood. A slight increase after examination of the experimental group, although not significant, suggests a positive effect of the intervention.

Internal Validity
Internal validity is essential in assessing the impact of online assessment on student performance. Our goal is to isolate other factors influencing the learning environment and to relate any observed changes in performance to changes in online assessment techniques.

External Validity:
Considering a broader context, we address external validity to ensure that our findings extend beyond the specific context of our research. In doing so, we seek to generalise our findings across educational settings, schools, and student populations.

Limitations and further research:

The study acknowledges several limitations, such as the short duration of the experiment and the possible influence of external factors not considered in the study design. Further research is to be conducted.